\documentclass{article}
\usepackage{eurosym}

\begin{document}

\begin{titlepage}
\title{Quantum cosmology with scalar fields:\\ self-adjointness and cosmological scenarios}

\author{Carla R. Almeida\footnote{carlagbjj@hotmail.com}, Antonio B. Batista\footnote{abrasilb918@gmail.com},
J\'ulio~C.~Fabris\footnote{fabris@pq.cnpq.br}\\
DF - UFES, Vit\'oria, ES, Brazil \vspace{0.5cm}\\
and\vspace{0.5cm}\\
Paulo R.L.V. Moniz\footnote{pmoniz@ubi.pt}\\
DF, Universidade da Beira Interior, Covilh\~a, Portugal\vspace{0.5cm}\\
}
\maketitle
\begin{abstract}
We discuss the issue of unitarity in particular quantum cosmological models with scalar field. The time variable is recovered, in this
context, by using the Schutz's formalism for a radiative fluid. Two cases are considered: a phantom scalar field and an ordinary scalar field. For the
first case, it is shown that the evolution is unitary provided a convenient factor ordering and inner product measure are chosen; the same happens for
the ordinary scalar field, except for some special cases for which the Hamiltonian is not self-adjoint but admits a self-adjoint extension. In all cases, even
for those cases not exhibiting unitary evolution, the formal computation of the expectation value of the scale factor indicates a non-singular bounce. The importance of the unitary evolution in quantum cosmology is briefly discussed.
\end{abstract}

\end{titlepage}

\section{Introduction}

Quantum cosmology faces difficulties that range from the technical to the
conceptual point of view. It is usually based on the ADM formulation of the General
Relativity theory \cite{adm1,adm2,adm3,nelsonr}, leading to the Wheeler-de
Witt equation, a functional equation defined in the superspace, the space of
all possible three-dimensional metrics on the space-like hypersurface which
foliates the four-dimensional space-time. Technically, the foremost problem is
to determine solutions of the Wheeler-de Witt equation in its
complete form, since it is a functional equation with an infinite number of
degrees of freedom. This suggests to freeze out an infinite number of such
degrees of freedom, reducing a system to a few numbers of variables, leading
to the mini-superspace setting. This drastic reduction implies that we ignore, for
example, the violation of the uncertainty principle to those degrees of
freedom that have been frozen out. On the other hand, the ADM formulation of
General Relativity forms a constrained system, and the Hamiltonian
constraints implies the absence of an explicit and clear time variable. On
the conceptual side, moreover, we must face the problem of a unique system,
the Universe at a whole, what leads to the inapplicability of the usual
Copenhagen interpretation of quantum mechanics.

The issue of the time variable has been addressed
in many different manners. One can, for example, identify an internal time, determining the evolution of the system, through the
deparametrisation procedure \cite{kuchar}. Or an external time (from the
point of view of the gravitational system) may be introduced, for example,
through the WKB approach (see, for example, reference \cite{colistete} and references therein) or
by considering matter fields. This last procedure has been used extensively
in the literature, and one possibility is to consider a fluid with internal
degrees of freedom using Schutz's description \cite
{schutz1,schutz2}. This approach leads to a Schr\"odinger-like equation,
since the momentum associated to the fluid variables appears linearly in the
resulting Hamiltonian \cite{rubakov}.

The interpretation problem is more delicate.  This very important issue will not be treated 
in a direct way in the present text, but the reader can address himself to many
text about this subject in the literature. See, as example, references \cite{nelsonr, tipler, bojowald,holland,nelson} and references
therein.

Our goal in this paper is to investigate another difficulty that appears even
when the mini-superspace approach is employed and the time variable is
identified through, for example, the Schutz's variable: the unitary evolution of resulting
quantum system. It has been shown that when just the Einstein-Hilbert
Lagrangian and a perfect fluid matter component is considered, the resulting
Hamiltonian can be made self-adjoint (unitarity is assured) under the
hypothesis of a maximally symmetric spatial section \cite{nivaldo}. However,
when anisotropy is taken into account, in the same context, the Hamiltonian
is generally no more self-adjoint \cite{brasilbis}. However, recently, it
has been shown that a convenient choice of the ordering factor associated
with the gravitational operators may restore the self-adjoint character of
the Hamiltonian \cite{pal}.

The breakdown of the unitary evolution when the time variable is
recovered through the Schutz variable occurs also when scalar fields are
considered together with the Einstein-Hilbert Lagrangian, even in the
isotropic e homogenous case \cite{brasil,moniz1}. Curiously, in this case,
and using the Schutz formalism to recover a Schr\"{o}dinger-like equation, a
phantom scalar field (with negative kinetic energy) assures the positivity
of the total energy, while for a \textit{normal} scalar field (with positive
kinetic energy) this positivity of the energy is not assured, since the
Hamiltonian has a hyperbolic signature for this case. It will be shown,
however, that a convenient factor ordering (similarly to what has been
discussed in the anisotropic case, reference \cite{pal}) together with a
convenient measure associated with the inner product guarantees a
self-adjoint Hamiltonian (hence a unitary evolution) for a phantom scalar
field. The same occurs for the normal scalar field, but for special cases
(connected with the ordering factor) the Hamiltonian is not self-adjoint but
admits a self-adjoint extension. We will not consider here the issue of unitary evolution
for other approaches to
recover the time variable as, for example, by using the scalar field itself as the time
coordinate \cite{vakili}.

This paper is organised as follows. In next section with consider a
scalar-tensor theory - specifically, the Brans-Dicke theory - coupled with a radiative fluid.
By performing conformal transformation, the theory is re-written in the
so-called Einstein's frame. In section 3, solutions for the
Schr\"odinger-like equation are obtained, and it is shown that they do not
display a unitary evolution. Even though, formal predictions for the evolution of
the Universe are obtained revealing a non-singular behaviour. In section 5
it is shown how to recover unitarity at least for the anomalous, phantom
configuration of the scalar field. The general self-adjoint issue is
considered in section 6. In section 7 we discuss the self-adjoint extension
for the particular cases where the Hamiltonian is not
self-adjoint but admits such an extension. In section 8 we present our conclusions.

\section{Scalar-tensor model with radiative fluid}

The prototype of the scalar-tensor gravity formulation is the Brans-Dicke
theory \cite{bd}, represented by the Lagrangian, 
\begin{eqnarray}  \label{classLD}
\mathcal{L} = \sqrt{-\tilde g}\phi\biggr\{\tilde R - \tilde \omega\frac{%
\phi_{;\rho}\phi^{;\rho}}{\phi^2}\biggl\} + \mathcal{L}_m,
\end{eqnarray}
where $L_m$ is the matter Lagrangian. We will suppose that the matter
Lagrangian refers to a radiative fluid, having conformal symmetry. This
Lagrangian includes as a particular case the string dilatonic Lagrangian for
which $\tilde\omega = - 1$ \cite{copeland}.

The Lagrangian (\ref{classLD}) defines the theory in the Jordan's frame.
Performing a conformal transformation such that $g_{\mu\nu} =
\phi^{-1}\tilde g_{\mu\nu}$, we transpose the action (\ref{classLD}) to the
corresponding expression written in the Einstein's frame \cite{frame}: 
\begin{eqnarray}  \label{class_L}
\mathcal{L} = \sqrt{-g}\biggr\{R - \omega\frac{\phi_{;\rho}\phi^{;\rho}}{%
\phi^2}\biggl\} + \mathcal{L}_m,
\end{eqnarray}
where $\omega = \tilde\omega + 3/2$. Remark that the matter Lagrangian is
not affected by the conformal transformation since we are considering a
radiative fluid.

We will consider from now on that the line element in the Einstein's frame
can be written as, 
\begin{equation}  \label{FLRW}
ds^2 = N(t)^2dt^2 - a(t)^2 \gamma_{ij}dx^i dx^j
\end{equation}
where $N(t)$ is the lapse function, $a(t)$ is the scale factor and $%
\gamma_{ij}$ is the induced metric of the homogeneous and isotropic spatial
hypersurfaces with curvature $k=0,\pm1$. For simplicity, we will fix $k = 0$%
. With this metric, the gravitational Lagrangian becomes, 
\begin{eqnarray}
\mathcal{L}_G = \frac{V_0a^3}{N}\biggr\{-6\biggr[\frac{\ddot a}{a} + \biggr(%
\frac{\dot a}{a}\biggl)^2 - \frac{\dot a}{a}\frac{\dot N}{N}\biggl] - \omega 
\frac{\dot \phi^2}{\phi^2}\biggl\}\quad ,
\end{eqnarray}
where $V_0$ is a constant and can be interpreted as the physical volume of
the compact universe (in this case a three-torus) divided by $a^3$. Since there is an identical
multiplicative constant in front of the matter Lagrangian we can drop it
from our analysis. Discarding a surface term, the gravitational Lagrangian
can be written as, 
\begin{eqnarray}
\mathcal{L}_G = \frac{1}{N}\biggr\{6a\dot a^2 - \omega a^3\frac{\dot \phi^2}{%
\phi^2} \biggl\}.
\end{eqnarray}
Defining, 
\begin{equation}
\sigma = \sqrt{|\omega|}\ln\phi,
\end{equation}
we obtain, 
\begin{eqnarray}
\mathcal{L}_G = \frac{1}{N}\biggr\{6a\dot a^2 - \epsilon a^3\dot\sigma^2 %
\biggl\},
\end{eqnarray}
where $\epsilon = \pm 1$ according $\omega$ is positive (upper sign) or
negative (lower sign). The canonical momenta associated with the scale
factor and the scalar field are respectively: 
\begin{eqnarray}  \label{momentum}
p_a = 12\frac{a\dot{a}}{N}\quad &,& \quad p_\sigma = - 2\epsilon \frac{%
a^3\dot\sigma}{N} \qquad,
\end{eqnarray}
leading to the following expression in terms of the conjugate momentum: 
\begin{eqnarray}
\mathcal{L}_G = p_a\dot{a}+p_\sigma\dot{\sigma}-N\biggr\{\frac{1}{24}\frac{%
p_a^2}{a} - \epsilon \frac{p_\sigma^2}{4a^3}\biggl\}\qquad.
\end{eqnarray}
Considering a radiative matter component (for the computation of the
conjugate momentum associated with the fluid, see references \cite%
{rubakov,nivaldo}), the total Hamiltonian is: 
\begin{eqnarray}
H = N\biggr\{\frac{1}{24}\frac{p_a^2}{a} - \epsilon\frac{p_\sigma^2}{4a^3}- 
\frac{p_T}{a}\biggl\}.
\end{eqnarray}
The resulting Schr\"odinger-like equation is, 
\begin{equation}  \label{se}
- \frac{\partial^2\Psi}{\partial a^2} + \frac{\epsilon}{a^2}\frac{%
\partial^2\Psi}{\partial\sigma^2} = i \frac{\partial\Psi}{\partial T},
\end{equation}
where we made the redefinition $\frac{\sigma}{\sqrt{6}} \rightarrow \sigma$
and $\frac{T}{24} \rightarrow T$.

\section{Cosmological scenarios}

In the reference\cite{moniz1}, the quantum cosmological model defined by the
Schr\"odinger-type equation (\ref{se}) has been studied. We will review in this section the procedure
and results obtained in reference \cite{moniz1}.
\par
In order to obtain
treatable expressions, the ordering ambiguity of operators has been
exploited, and equation (\ref{se}) has been rewritten as, 
\begin{equation}  \label{sebis}
- \frac{\partial^2\Psi}{\partial a^2} - \frac{1}{a}\partial_a\Psi + \frac{%
\epsilon}{a^2}\frac{\partial^2\Psi}{\partial\sigma^2} = i \frac{\partial\Psi%
}{\partial T},
\end{equation}
For $\epsilon = - 1$, a condition that assures the positivity of energy,
equation (\ref{sebis}) admits a solution in terms of stationary states of
energy $E$: 
\begin{equation}  \label{sol}
\Psi = AJ_\nu(\sqrt{E}a)e^{ik\sigma}e^{-iET}, \quad \nu = k\quad ,
\end{equation}
with $A$ being a normalization constant and $k$ is a separation constant.

A particular wavepacket can be obtained by a convenient superposition of the constants $%
E$ and $k$, as it is described in \cite{moniz1}. The final result is 
\begin{equation}  \label{wf1}
\Psi(a,\sigma,T) = C\frac{e^{- \frac{a^2}{4B(T)}}}{B(T)\,g_\alpha(a,B,\sigma)%
},
\end{equation}
where 
\begin{equation}
B(T) = (\gamma + i\,T), \quad g_\alpha(a,\sigma,T) = - \alpha + \,\ln\biggr[%
\frac{a}{2B(T)}\biggl] \pm i\sigma,
\end{equation}
$\gamma$ and $\alpha$ being constants connected with the gaussian-type superposition, and $C$ is a normalisation constant.

The norm of the wavefunction (\ref{wf1}) can be calculated explicitly: 
\begin{eqnarray}
N &=& \int_0^\infty\int_{-\infty}^{+\infty}\Psi^*\Psi\,da\,d\sigma =\frac{C^2%
}{(B\,B^*)^{1/2}} \pi\int_0^\infty \frac{e^{-\gamma u^2}}{\alpha + \ln\biggr(%
\frac{u}{2}\biggl)}du  \nonumber \\
&=&\frac{C^2}{(B\,B^*)^{1/2}} \pi I_1 ,
\end{eqnarray}
where $I_1$ is the definite integral, 
\begin{equation}
I_1 = \int_0^\infty \frac{e^{-\gamma u^2}}{\alpha + \ln\biggr(\frac{u}{2}%
\biggl)}du.
\end{equation}
The norm is clearly time-dependent: the corresponding quantum model is not
unitary. Even though, the expectation value for the scalar field can be formally
computed, leading to the expression,
\begin{eqnarray}
<a> \propto (\gamma^2 + T^2)^{1/2}.
\end{eqnarray}
The same result is, essentially, obtained through the computation of the 
bohmian trajectories. The expectation value for the scale factor indicates
a non-singular bounce. As it is shown in reference \cite{moniz1}, the expectation value of the
scalar field
$\sigma$ is time-dependent, reading
\begin{eqnarray}
<\sigma> \propto \arctan\biggr(\frac{T}{\gamma}\biggl).
\end{eqnarray}

\section{Recovering unitarity}

Let us suppress the ordering factor introduced previously. The Schr\"odinger
equation is 
\begin{eqnarray}
-\partial_a^2\Psi + \frac{\epsilon}{a^2}\partial_\sigma^2\Psi =
i\partial_T\Psi,
\end{eqnarray}
with $\epsilon = \pm 1$. For a stationary state, 
\begin{eqnarray}
\Psi = \phi e^{-iET},
\end{eqnarray}
leading to 
\begin{eqnarray}
-\partial_a^2\phi + \frac{\epsilon}{a^2}\partial_\sigma^2\phi = E\phi.
\end{eqnarray}

Using the separation variables method, such that 
\begin{eqnarray}
\phi(a,\sigma) = X(a)Y(\sigma),
\end{eqnarray}
the equation becomes, 
\begin{eqnarray}
- \frac{X^{\prime \prime }}{X} + \frac{\epsilon}{a^2}\frac{\ddot Y}{Y} = E,
\end{eqnarray}
where the primes mean derivative with respect to $a$ and the dots mean
derivative with respect to $\sigma$. This equation can be rewritten as, 
\begin{eqnarray}
\epsilon\frac{\ddot Y}{Y} = a^2\biggr(E + \frac{X^{\prime \prime }}{X}%
\biggl) = - \epsilon k^2,
\end{eqnarray}
where $k$ is a constant of separation. In this case, the function $Y$
satisfies the equation, 
\begin{eqnarray}
\ddot Y + k^2Y = 0,
\end{eqnarray}
with the solution, 
\begin{eqnarray}
Y = Y_0 e^{ik\sigma}.
\end{eqnarray}

The equation for $X$ reads, 
\begin{eqnarray}
X^{\prime \prime }+ \biggr(E + \epsilon\frac{k^2}{a^2}\biggl)X = 0.
\end{eqnarray}
The soluton is, 
\begin{eqnarray}
X(a) = \sqrt{a}J_\nu(\sqrt{E}a), \quad \nu = \sqrt{\frac{1}{4} - \epsilon k^2%
}.
\end{eqnarray}
The total wavefunction is, 
\begin{eqnarray}
\Psi = \Psi_0\sqrt{a}J_\nu(\sqrt{E}a)e^{-i(k\sigma + ET)}.
\end{eqnarray}

Let us construct the wavepacket. First we write $x=\sqrt{E}$. Then, we can
made the following superposition: 
\begin{eqnarray}
\Psi (a,\sigma ) = \sqrt{a}\int_{-\infty }^{+\infty }\int_{0}^{+\infty
}A(k)x^{\nu +1}e^{- Bx^{2}}J_{\nu }(xa)e^{-ik\sigma }dk\,dx, 
\end{eqnarray}
where, as before,
\begin{eqnarray}
B = \gamma + iT, 
\end{eqnarray}
$\gamma $ being a positive constant (in order to assure the convergence of
the integral), playing the same r\^ole as the previous section. The function $A(k)$ will remain for the moment unspecified,
but it must decay exponentially for large $k$ in order also to assure the
convergence of the wavepacket. The integral in $x$ can be performed leading
to \cite{grad}, 
\begin{eqnarray}
\label{wps}
\Psi =\int_{-\infty }^{+\infty }A(k)e^{-ik\sigma}\frac{a^{\nu +1/2}}{(2B
)^{\nu +1}}\exp \biggr(-\frac{a^{2}}{4B }\biggl)dk. 
\end{eqnarray}

From now on, let us consider the case $\epsilon =-1$, such that $\nu $ is
real. The modulus of the wavefunction reads, 
\begin{eqnarray}
\Psi ^{\ast }\Psi =\int_{-\infty}^{+\infty} \int_{-\infty}^{+\infty} A(k)A^{\ast }(k^{\prime })e^{-i(k-k^{\prime
})\sigma }B^{-\nu -1}{B^{\ast }}^{-\nu ^{\prime }-1}a^{\nu +\nu
^{\prime }+1}\nonumber\\
\times\exp\biggr[-\frac{\gamma a^{2}}{2\bar B}\biggl]dkdk^{\prime }, 
\end{eqnarray}
where 
$\bar B = B B^{\ast } = \gamma ^{2}+T^{2}$.

The integration in $\sigma $ in the interval $-\infty <\sigma <+\infty $
implies a delta function, $\delta (k-k^{\prime })$. After integrating in $
k^{\prime }$, we find 
\begin{eqnarray}
\int_{-\infty }^{+\infty }\Psi ^{\ast }\Psi d\sigma =\int_{-\infty
}^{+\infty }|A(k)|^{2}\bar B^{-2\nu -1}a^{2\nu +1}e^{-\frac{a^{2}}{2B}}\,dk. 
\end{eqnarray}
Performing the change of variable, 
\[
y=\frac{a}{\sqrt{\bar B}}, 
\]%
we can write the norm as, 
\begin{eqnarray}
N = \int_0^{+\infty}\int_{-\infty }^{+\infty }|A(k)|^{2}e^{-\frac{\gamma }{2}y^{2}}y^{2\nu
+1}\,dy\,dk. 
\end{eqnarray}
The norm is time-independent. After integrating in $y$, it can also be
written as, 
\begin{eqnarray}
N=\frac{1}{2}\int_{-\infty }^{+\infty }|A(k)|^{2}\Gamma \biggr(\frac{3+2\nu 
}{4}\biggl)\,dk. 
\end{eqnarray}
Due to the asymptotic properties of the $\Gamma $'s function, the integral
converges if $A(k)$ behaves, for example, as $e^{-k^{2}}$ for large values
of $k$.

The expectation value for the scalar field is given by the expression,
\begin{eqnarray}
<a> = \frac{1}{N}\int_{-\infty}^{+\infty}\int_0^{+\infty}\Psi^* a \Psi\,d\sigma\,da.
\end{eqnarray}
Using the same procedure as before, we end up with,
\begin{eqnarray}
<a> = \frac{1}{N}\sqrt{\bar B}\int_{-\infty}^{+\infty}\int_0^{+\infty}|A(k)|^2\,e^{-\frac{\gamma}{2}y^2}\, y^{2(\nu + 1)}\,dy\,dk \propto (\gamma^2 + T^2)^{1/2},
\end{eqnarray}
which is essentially the same prediction as in the previous non-unitary case.

We can also evaluate the expectation value for the scalar
field. In fact, to obtain the expectation value of $\sigma$ we must compute
\begin{eqnarray}
<\sigma> \,\propto\, \int_0^\infty\int_{-\infty}^{+\infty}\Psi(a,\sigma)^*\sigma\Psi(a,\sigma)da\,d\sigma.
\end{eqnarray}
Using (\ref{wps}), we can write
\begin{eqnarray}
\sigma\Psi = i\int_{-\infty }^{+\infty }A(k)[\partial_k e^{-ik\sigma}]\frac{a^{\nu +1/2}}{(2B
)^{\nu +1}}\exp \biggr(-\frac{a^{2}}{4B }\biggl)dk,
\end{eqnarray} 
which can be integrated by parts. Using this expression, the integration in $\sigma$ leads again to delta function. Redefining, as before, the integration in $a$, we obtain the following expression,
\begin{eqnarray}
<\sigma>_T &=& - 2\pi i\int_{-\infty}^{+\infty}\int_0^{+\infty} |A(k)|^2\,e^{- \frac{\gamma}{2}y^2}\biggr\{\frac{A_k(k)}{A(k)} \nonumber\\
&+& \nu_k\ln\biggr(\frac{y}{2}\biggl)
- \nu_k\biggr(\ln\sqrt{\frac{B^*}{B}}\biggl)\biggl\}\,\biggr(\frac{y}{2}\biggl)^{2\nu + 1}\,dy\,dk,
\end{eqnarray}
where the subscript $k$ indicates derivative with respect to this parameter.
After expressing $B$ in polar form, we obtain,
\begin{eqnarray}
<\sigma>_T = \sigma_0 + \sigma_1\arctan\biggr(\frac{T}{\gamma}\biggl),
\end{eqnarray}
where $\sigma_0$ and $\sigma_1$ are constants.
This is essentially the same result obtained in the previous section for the expectation value of $\sigma$.
Remark, however, that $<\sigma> = 0$ if $A(k)$ is an even function, such that $A(k) = A(-k)$.

\section{Self-adjointness}

The results shown before reveal that the self-adjointness property of the
effective Hamiltonian may depend on the ordering factor. The overall
situation is, however, more involved. The Hamiltonian used in equation (\ref%
{sebis}) is, 
\begin{eqnarray}  \label{h1}
H = - \partial_a^2 - \frac{1}{a}\partial_a + \frac{\epsilon}{a^2}%
\partial_\sigma^2.
\end{eqnarray}
When $\epsilon = - 1$, this operator looks like that one of the two
dimensional problem written in polar coordinates. In fact, if we make the
identification $a \equiv r$, $\sigma \equiv \theta$, considering the
variable $\theta$ as periodic, such that we can define the cartesian
coordinates $x = r\cos\theta$ and $y = r\sin\theta$, the Hamiltonian takes
the form, 
\begin{eqnarray}
H = - \partial^2_x - \partial_y^2,
\end{eqnarray}
with $- \infty < x, y < + \infty$, which is clearly self-adjoint.

What makes the Hamiltonian (\ref{h1}) not self-adjoint? We could think that
is the fact that the coordinate $\sigma$ there is not periodic. But, we
remark that in passing from cartesian to polar coordinates, there is a
Jacobian factor that it has not been used in computing the inner product in
the Hilbert space for the Hamiltonian (\ref{h1}). The clue to this problem
seems to be in the ordering factor and in the measure of the inner product,
as point out in reference \cite{pal} in the context of anisotropic models.

Let us be more specific. Let us consider the Hamiltonian with a ordering
factor labeled by $q$: 
\begin{eqnarray}  \label{hf}
\hat H = - \partial_a^2 - \frac{q}{a}\partial_a + \frac{\epsilon}{a^2}%
\partial_\sigma^2,
\end{eqnarray}
where we have restored the factor $\epsilon = \pm 1$. This Hamiltonian is
symmetric under the inner product defined by, 
\begin{eqnarray}
(\phi,\Psi) =
\int_0^\infty\int_{-\infty}^{+\infty}\phi^*\Psi\,a^q\,da\,d\sigma,
\end{eqnarray}
if the functions $\phi$ and $\Psi$, as well as their first derivatives, are
null in the extreme of the interval.

To be symmetric is a necessary but not sufficient condition for the operator 
$\hat H$ to be self-adjoint: the domain of the operator and of is hermitian
conjugate must be also the same. The self-adjoint character of a symmetric
Hamiltonian can be obtained through the deficiencies indices of von Neumann 
\cite{reed,griffiths}.

The deficiencies indices are defined as follows. Consider the eigenvalue
problem, 
\begin{eqnarray}  \label{vn}
\hat H\Psi = \pm i\Psi.
\end{eqnarray}
Let us call $n_+$ and $n_-$ the number of square integrable solutions of the
eigenvalue problem given by equation (\ref{vn}) when the upper and lower
sign in the right hand side is chosen, respectively. If $n_+ = n_- = 0$, the
operator $\hat H$ is already self-adjoint; if $n_+ = n_- \neq 0$, the
operator is not self-adjoint but it admits self-adjoint extensions given by
some restrictions in the wavefunctions; if $n_+ \neq n_-$ the operator is
not self-adjoint and it does not admit any self-ajoint extension.

Hence, we must solve the eigenvalue equation, 
\begin{eqnarray}
-\partial _{a}^{2}\Psi -\frac{q}{a}\partial _{a}\Psi +\frac{\epsilon }{a^{2}}%
\Psi =\pm i\Psi =\eta \Psi , 
\end{eqnarray}
where we have defined $\eta =\pm i$. The solutions of this equation can be
written as 
\begin{eqnarray}
\Psi =a^{p}\biggr[H_{\nu }^{(1)}(\sqrt{\eta }a)+H_{\nu }^{(2)}(\sqrt{\eta }a)%
\biggl]e^{\pm ik\sigma }, 
\end{eqnarray}
where, 
\begin{eqnarray}
p=\frac{1-q}{2},\quad \nu =\sqrt{p^{2}-\epsilon k^{2}}, 
\end{eqnarray}
$k$ is a separation constant which must be real (otherwise, in the
corresponding Schr\"{o}dinger equation, we would have just non-integrable
solutions, with divergent norm), and $H_{\nu }^{(1,2)}(x)$ are Hankel's
functions of first and second kind.

Let us write the functions, 
\begin{eqnarray}
\Psi _{\pm }^{(1)} &=&a^{p}\,H_{\nu }^{(1)}(\eta a)e^{ik\sigma }, \\
\Psi _{\pm }^{(2)} &=&a^{p}\,H_{\nu }^{(2)}(\eta a)e^{ik\sigma }.
\end{eqnarray}
These solutions contain the separation parameter $k$, which parametrises the
plane wave behaviour in terms of the coordinate $\sigma $. In order to give
a physical meaning to such structure is necessary to construct a wave packet
related to these plane wave solutions. This procedure has the purpose to
avoid unphysical behaviour, but it has no major impact on the considerations
on the self-adjoint character of the operator. Hence, the solutions take the
form, 
\begin{eqnarray}
\Psi _{\pm }^{(1)} &=&\int_{-\infty }^{+\infty }A(k)\,a^{p}\,H_{\nu
}^{(1)}(\eta a)e^{ik\sigma }\,dk,  \label{wf2} \\
\Psi _{\pm }^{(2)} &=&\int_{-\infty }^{+\infty }A(k)\,a^{p}\,H_{\nu
}^{(2)}(\eta a)e^{ik\sigma }\,dk,  \label{wf3}
\end{eqnarray}%
where $A(k)$, as before, is a superposition factor which satisfies the condition to go
to zero sufficiently fast at both infinities.

We must investigate the norm of the wave functions (\ref{wf2},\ref{wf3}),
given by 
\begin{eqnarray}
N_{\pm }^{(1,2)}=\int_{-\infty }^{+\infty }\int_{0}^{+\infty }\Psi _{\pm
}^{(1,2)}(a,\sigma )\,{\Psi _{\pm }^{(1,2)}}^{\ast }(a,\sigma
)\,a^{q}\,da\,d\sigma . 
\end{eqnarray}
First, let us consider the integration in $\sigma $. We have: 
\begin{eqnarray}
I_{\sigma } &=&\int_{-\infty }^{+\infty }{\Psi _{\pm }^{(1,2)}}^{\ast
}\,\Psi _{\pm }^{(1,2)}\,d\sigma  \nonumber \\
&=&\int_{-\infty }^{+\infty }\int_{-\infty }^{+\infty }\int_{-\infty
}^{+\infty }A(k)A^{\ast }(\bar{k})a^{2p}\,{H_{\bar{\nu}}^{(1,2)}(\eta a)}%
^{\ast }H_{\nu }^{(1,2)}(\eta a)\,e^{i(k-\bar{k})\sigma }\,dk\,d\bar{k}%
\,d\sigma  \nonumber \\
&=&2\pi \int_{-\infty }^{+\infty }\int_{-\infty }^{+\infty }A(k)A^{\ast }(%
\bar{k})a^{2p}\,{H_{\bar{\nu}}^{(1,2)}}^{\ast }(\eta a)H_{\nu }^{(1,2)}(\eta
a)\,\delta (k-\bar{k})\,dk\,d\bar{k}  \nonumber \\
&=&2\pi \int_{-\infty }^{+\infty }|A(k)|^{2}a^{2p}\,{H_{\bar{\nu}}^{(1,2)}}%
^{\ast }(\eta a)H_{\nu }^{(1,2)}(\eta a)\,dk.
\end{eqnarray}%
The norm becomes, 
\begin{eqnarray}
N &=&\int_{-\infty }^{+\infty }\int_{0}^{+\infty }{\Psi _{\pm }^{(1,2)}}%
^{\ast }\Psi _{\pm }^{(1,2)}\,a^{q}\,da\,d\sigma  \nonumber \\
&=&2\pi \int_{0}^{+\infty }\int_{-\infty }^{+\infty }|A(k)|^{2}a^{2p+q}\,{H_{%
\bar{\nu}}^{(1,2)}}^{\ast }(\eta a)H_{\nu }^{(1,2)}(\eta a)\,dk\,da 
\nonumber \\
&=&2\pi \int_{0}^{+\infty }\int_{-\infty }^{+\infty }|A(k)|^{2}a\,\,{H_{\bar{%
\nu}}^{(1,2)}}^{\ast }(\eta a)H_{\nu }^{(1,2)}(\eta a)\,dk\,da.
\end{eqnarray}

We may investigate separately the norm in two limits, when $a\rightarrow 0$
and $a\rightarrow \infty $.

\begin{itemize}
\item $a\rightarrow 0$. In this case, 
\begin{equation}
H_{\nu }^{(1,2)}(\eta a)\sim a^{-\nu }.  \label{eq. a-->0}
\end{equation}
The norm behaves as, 
\begin{eqnarray}
N \sim \int_{-\infty}^{+\infty} |A(k)|^2a^{2[1-\nu(k)]}dk.,
\end{eqnarray}
where we have clearly indicated that $\nu$ depends on $k$.
We have two different situations.

\begin{enumerate}
\item If $\epsilon = - 1$, $\nu \in \Re^+$. In fact, $|p| \leq \nu < \infty$. There is no square-integrable eigenfunction in this
case. This indicates that the Hamiltonian operator is self-adjoint.

\item If $\epsilon = 1$, $\nu = \sqrt{p ^2 - k^2}$, and the divergence of
the norm in this limit depends on the ordering factor and on the interval of 
$k$. Supposing, for example, that the superposition factor $A(k)$ is defined
in all interval of $k$, there is always divergences unless $- \sqrt{1 - p^2} \leq k \leq + \sqrt{1 - p^2}$. In order to assure the existence of this interval of values, it is required that $p \leq 1$. 
\end{enumerate}

\item $a\rightarrow \infty $. In this limit, 
\begin{eqnarray}
H_{\nu }^{(1,2)}(\eta a)\sim \sqrt{\frac{2}{\pi \sqrt{\eta }a}}e^{\pm i(
\sqrt{\eta }a-\frac{\pi \nu }{2}-\frac{\pi }{4})}, 
\end{eqnarray}
where the upper (lower sign) in the exponential corresponds to the Hankel's
function of the first (second) kind. In this limit, the norm of the wave
functions takes the form, 
\[
N_{\pm }\sim \frac{2}{\pi }\int_{0}^{+\infty }\int_{-\infty }^{+\infty
}|A(k)|^{2}\,a\,e^{\pm i[(\pm )i\,a-\frac{\pi }{2}(\nu -\nu ^{\ast
})]}dk\,da, 
\]%
where the symbol $(\pm )$ corresponds to the two types of eigenfunctions of
the equation (\ref{vn}). It is clear that for each type of Hankel's function
there is one eigenfunction that converges (leading to a finite contribution to the norm) and the other that diverges (leading to a infinite contribution to the norm).
\end{itemize}

Hence, with these informations for the two asymptotic regimes, we can
conclude the following.

\begin{enumerate}
\item For $\epsilon = - 1$ we have just divergent eigenfunctions
corresponding to the equation ($\ref{vn}$) due to the behaviour in the limit 
$a \rightarrow 0$. This implies the deficiency indices are such that $n_+ =
n_- = 0$. The Hamiltonian operator is self-adjoint.

\item For $\epsilon =1$, there are two possibilities.

\begin{enumerate}
\item For an ordering factor such that $p > 1$, we find, as previously,
divergent eigenfunctions for equation (\ref{vn}) implying $n_+ = n_- = 0$
and the Hamiltonian operator is already self-adjoint.

\item For an ordering factor such that $p \leq 1$, the wave functions are
convergent in the limit $a\rightarrow 0$, but in the limit $a\rightarrow
\infty $ we have one divergent and one convergent function for each sign in
the equation (\ref{vn}). These asymptotic behaviours imply $n_{+}=n_{-}=1$.
Hence, the Hamiltonian is not self-adjoint but admits a self-adjoint
extension.
\end{enumerate}
\end{enumerate}

\section{Self-adjoint extension}

The conclusion of the preceding section was that the case $\epsilon = - 1$
leads to a Hamiltonian that is already self-adjoint, but for the case $%
\epsilon = 1$ the Hamiltonian is self-adjoint if the factoring order $p$ is
greater than 1, while for $p \leq 1$ the Hamiltonian is not self-adjoint,
admitting however a self-adjoint extension. The goal of the present section
is to verify the conditions on the wave functions in order to have this
self-adjoint extension when $\epsilon = 1$ and $p \leq 1$.

The self-adjoint extention are given by the functions $\phi $ such that, 
\begin{eqnarray}
\left\langle \psi |H\phi \right\rangle =\left\langle H\psi |\phi
\right\rangle , 
\end{eqnarray}
where $\psi =\Psi _{+}+\lambda \Psi _{-}$ for some $\lambda $. We will keep,
just for generality, the value of $\epsilon $ undefined. We have then, 
\begin{eqnarray}
\left\langle \psi |H\phi \right\rangle &=&\int_{-\infty }^{\infty
}\int_{0}^{\infty }\psi ^{\ast }\left( H\phi \right) a^{q}dad\sigma 
\nonumber \\
&=&\int \int \psi ^{\ast }\left( -\partial _{a}^{2}\phi -\frac{q}{a}\partial
_{a}\phi +\frac{\epsilon }{a^{2}}\partial _{\sigma }^{2}\phi \right)
a^{q}dad\sigma .
\end{eqnarray}%
Let's compute the integral each term separately. The first terms read,
\begin{eqnarray}
&&\int \int -\psi ^{\ast }\left( \partial _{a}^{2}\phi \right)
a^{q}dad\sigma =\int_{-\infty }^{\infty }\left[ -\psi ^{\ast }a^{q}\partial
_{a}\phi \right] _{a=0}^{a=\infty }-\left[ -\partial _{a}\left( \psi ^{\ast
}a^{q}\right) \phi \right] _{a=0}^{a=\infty }d\sigma  \nonumber \\
&&+\int_{-\infty }^{\infty }\int_{0}^{\infty }-\partial _{a}^{2}\left( \psi
^{\ast }a^{q}\right) \phi dad\sigma  \nonumber \\
&=&\int_{-\infty }^{\infty }\left\{ \left[ \left( \partial _{a}\psi ^{\ast
}\right) \phi -\psi ^{\ast }\left( \partial _{a}\phi \right) \right]
a^{q}+qa^{q-1}\psi ^{\ast }\phi \right\} _{a=0}^{a=\infty }d\sigma  \nonumber\\
&&+\int\int -\left( \partial _{a}^{2}\psi ^{\ast }\right) \phi
a^{q}-2q\left( \partial _{a}\psi ^{\ast }\right) \phi a^{q-1}-q\left(
q-1\right) \psi ^{\ast }\phi a^{q-2}dad\sigma.
\end{eqnarray}
The second term is given by,
\begin{eqnarray}
&&\int \int -\psi ^{\ast }\left( \frac{q}{a}\partial _{a}\phi \right)
a^{q}dad\sigma =\int_{-\infty }^{\infty }\left[ -q\psi ^{\ast }\phi a^{q-1}%
\right] _{a=0}^{a=\infty }d\sigma  \nonumber \\
&&+\int_{-\infty }^{\infty }\int_{0}^{\infty }q\left( \partial _{a}\psi
^{\ast }\right) \phi a^{q-1}+q\left( q-1\right) \psi ^{\ast }\phi
a^{q-2}dad\sigma ,
\end{eqnarray}
Finally, the third term is, 
\begin{eqnarray}
&&\int_{-\infty }^{\infty }\int_{0}^{\infty }\psi ^{\ast }\left( \frac{%
\epsilon }{a^{2}}\partial _{\sigma }^{2}\phi \right) a^{q}dad\sigma
=\int_{0}^{\infty }\frac{\epsilon }{a^{2}}\left[ \psi ^{\ast }\left(
\partial _{\sigma }\phi \right) -\left( \partial _{\sigma }\psi ^{\ast
}\right) \phi \right] _{\sigma =-\infty }^{\sigma =\infty }a^{q}da  \nonumber
\\
&&+\int_{-\infty }^{\infty }\int_{0}^{\infty }\frac{\epsilon }{a^{2}}\left(
\partial _{\sigma }^{2}\psi ^{\ast }\right) \phi a^{q}dad\sigma .
\end{eqnarray}%
Putting it all those expressions together, we obtain 
\begin{eqnarray}
&&\left\langle \psi |H\phi \right\rangle =\int_{-\infty }^{\infty }\left\{ 
\left[ \left( \partial _{a}\psi ^{\ast }\right) \phi -\psi ^{\ast }\left(
\partial _{a}\phi \right) \right] a^{q}\right\} _{a=0}^{a=\infty }d\sigma 
\nonumber \\
&+&\int_{0}^{\infty }\frac{\epsilon }{a^{2}}\left[ \psi ^{\ast }\left(
\partial _{\sigma }\phi \right) -\left( \partial _{\sigma }\psi ^{\ast
}\right) \phi \right] _{\sigma =-\infty }^{\sigma =\infty }a^{q}da  \nonumber
\\
&&+\int \int \left( -\partial _{a}^{2}\psi ^{\ast }-\frac{q}{a}\partial
_{a}\psi ^{\ast }+\frac{\epsilon }{a^{2}}\partial _{\sigma }^{2}\psi ^{\ast
}\right) \phi a^{q}dad\sigma .
\end{eqnarray}%
Then, in order to get $\left\langle \psi |H\phi \right\rangle =\left\langle
H\psi |\phi \right\rangle $, the function $\phi $\ must satisfy the boundary
conditions: 
\begin{eqnarray}
\left\{ \left[ \left( \partial _{a}\psi ^{\ast }\right) \phi -\psi ^{\ast
}\left( \partial _{a}\phi \right) \right] a^{q}\right\} _{a=0}^{a=\infty }
&=&0; \\
\left[ \psi ^{\ast }\left( \partial _{\sigma }\phi \right) -\left( \partial
_{\sigma }\psi ^{\ast }\right) \phi \right] _{\sigma =-\infty }^{\sigma
=\infty } &=&0.
\end{eqnarray}%
Nevertheless, since $\phi $ is already square integrable, and goes to zero
sufficiently quickly in the extremes of the interval of $\sigma $, and the
same happens for $a\rightarrow \infty $, the only condition that is not
automatically satisfied is 
\begin{eqnarray}
\lim_{a\rightarrow 0}\left\{ \left[ \left( \partial _{a}\psi ^{\ast }\right)
\phi -\psi ^{\ast }\left( \partial _{a}\phi \right) \right] a^{q}\right\}
=0. 
\end{eqnarray}

Remembering that we found previously, 
\begin{eqnarray*}
\Psi _{+} &=&\int_{-\infty }^{\infty }A\left( k\right) a^{p}H_{\nu }^{\left(
1\right) }\left( \sqrt{i}a\right) e^{ik\sigma }dk \\
\Psi _{-} &=&\int_{-\infty }^{\infty }A\left( k\right) a^{p}H_{\nu }^{\left(
2\right) }\left( \sqrt{-i}a\right) e^{ik\sigma }dk,
\end{eqnarray*}%
then, 
\begin{eqnarray}
\psi &=&\int_{-\infty }^{\infty }A\left( k\right) a^{p}\left[ H_{\nu
}^{\left( 1\right) }\left( \sqrt{i}a\right) +\lambda H_{\nu }^{\left(
2\right) }\left( \sqrt{-i}a\right) \right] e^{ik\sigma }dk; \\
\psi ^{\ast } &=&\int_{-\infty }^{\infty }A^{\ast }\left( k\right) a^{p} 
\left[ H_{\nu }^{\left( 1\right) }\left( \sqrt{i}a\right) +\lambda H_{\nu
}^{\left( 2\right) }\left( \sqrt{-i}a\right) \right] ^{\ast }e^{-ik\sigma
}dk.
\end{eqnarray}%
The condition is given for $a\longrightarrow 0$ and, in this limite we have
the expression (\ref{eq. a-->0}). Therefore 
\[
\psi ^{\ast }\sim \int_{-\infty }^{\infty }A^{\ast }\left( k\right)
a^{p}\left( a^{-\nu }+\lambda a^{-\nu }\right) e^{-ik\sigma }dk. 
\]%
Hence, we must have, 
\begin{eqnarray}
&&\lim_{a\rightarrow 0}\partial _{a}\left[ \int_{-\infty }^{\infty }A^{\ast
}\left( k\right) a^{p}\left( a^{-\nu }+\lambda a^{-\nu }\right) e^{-ik\sigma
}dk\right] \phi a^{q}  \nonumber \\
&&-\lim_{a\rightarrow 0}\left[ \int_{-\infty }^{\infty }A^{\ast }\left(
k\right) a^{p}\left( a^{-\nu }+\lambda a^{-\nu }\right) e^{-ik\sigma }dk%
\right] \left( \partial _{a}\phi \right) a^{q}  \nonumber \\
&=&\lim_{a\rightarrow 0}\left[ \int_{-\infty }^{\infty }A^{\ast }\left(
k\right) \left( p-\nu \right) a^{p-\nu -1}e^{-ik\sigma }+A^{\ast }\left(
k\right) \lambda \left( p-\nu \right) a^{p-\nu -1}e^{-ik\sigma }dk\right]
\phi a^{q}  \nonumber \\
&&-\lim_{a\rightarrow 0}\left[ \int_{-\infty }^{\infty }A^{\ast }\left(
k\right) a^{p}\left( a^{-\nu }+\lambda a^{-\nu }\right) e^{-ik\sigma }dk%
\right] \left( \partial _{a}\phi \right) a^{q}  \nonumber \\
&=&\lim_{a\rightarrow 0}\left[ \frac{\left( p-\nu \right) }{a}\int_{-\infty
}^{\infty }A^{\ast }\left( k\right) a^{p}\left( a^{-\nu }+\lambda a^{-\nu
}\right) e^{-ik\sigma }dk\right] \phi a^{q}  \nonumber \\
&&-\lim_{a\rightarrow 0}\left[ \int_{-\infty }^{\infty }A^{\ast }\left(
k\right) a^{p}\left( a^{-\nu }+\lambda a^{-\nu }\right) e^{-ik\sigma }dk%
\right] \left( \partial _{a}\phi \right) a^{q}  \nonumber \\
&=&\lim_{a\rightarrow 0}\left[ \int_{-\infty }^{\infty }A^{\ast }\left(
k\right) a^{p+q}\left( a^{-\nu }+\lambda a^{-\nu }\right) e^{-ik\sigma }dk%
\right] \left[ \frac{\left( p-\nu \right) \phi }{a}-\left( \partial _{a}\phi
\right) \right]
\end{eqnarray}%
Since $p+q=p+(-2p-1)=-p-1<0$ and $\nu >0$, the integral term diverges and
the condition reduces to 
\begin{eqnarray}
\label{cond}
\lim_{a\rightarrow 0}\left[ \frac{\left( p-\nu \right) \phi }{a}-\left(
\partial _{a}\phi \right) \right] =0. 
\end{eqnarray}
Then, the self-adjoint extension is given in the subset of the square
integrable function that satisfies the condition (\ref{cond}).

\section{Conclusions}

In this paper we have addressed the problem of unitary evolution of
isotropic and homogenous quantum cosmological model in scalar-tensor
theories of gravity, with a time variable defined by the matter fluid. As it
happens in the case of the anisotropic universe without scalar field \cite{pal}, the unitarity is
assured if an ordering factor and an inner product measure are conveniently
chosen. We have considered the case where the matter fluid is radiation,
hence, bearing conformal invariance. This is somehow crucial in our analysis since
we have consider initially a scalar-tensor theory defined in the Jordan's
frame, and transformed it to a minimal coupling configuration, the
Einstein's frame. Hence, our conclusions are restricted to such framework.

The analysis we have performed uses the traditional techniques of quantum
mechanics, including the computation of the deficiency indices to determine
the self-adjoint character of the Hamiltonian, as well as the verification
of the self-adjoint extension for the case the Hamiltonian operator is not
self-adjoint but admits a self-adjoint extension \cite{reed,griffiths}. 

There is an essential point we have touched only superficially in the present work: given a
self-adjoint operator, that assures a unitary evolution for the system,
what are the predictions to evolution of the Universe? In reference \cite
{moniz1}, where a non-unitary model has been considered, the predictions for
the evolution of the universe were obtained by considering the bohmian's
trajectories, and a non-singular universe, with a bounce, has been obtained.
After restoring unitarity, at least for a particular case, essentially the same predictions 
were obtained either for the scale factor and for the scalar field.

The results described above may lead to a
different issues:  what is the r\^ole of unitary evolution in the universe and how it is connected with
the choice of given time coordinate. In
references \cite{nelson1,nelson2}, for example, scalar-tensor theories were
analysed without introducing an external time variable connected with a
dynamical fluid, and the predictions obtained through the WKB method and by bohmian trajectories indicated a non-trivial behaviour for
the scalar field and for the scale factor. It is important to compare these different procedures.

The r\^ole played by the unitary evolution, which supposes the choice of a
variable parametrising the evolution of the system, and the equivalence
between different choices for this time variable, must be discussed
more deeply. 
The consequences of not having a positive defined energy when the
scalar field has positive kinetic energy (but with a self-adjoint Hamiltonian) must be also investigated.
We hope to address these questions in future works.
 \newline
\vspace{0.3cm} \newline
\textbf{Acknowledgements}. JCF thanks CNPq (Brazil) and FAPES (Brazil) for partial
financial support. CRA thanks CAPES (Brazil) for financial support. We thank Giuseppe Dito and
Nelson Pinto-Neto for many enlightening discussions.

\end{document}